\documentclass[aps,preprint,showpacs]{revtex4}
\usepackage{graphics}
\usepackage{graphicx}

\def\bomega{\mbox{\boldmath $\omega$}}
\def\ss{\scriptscriptstyle }
\begin{document}
\title{Transient magnetoconductivity of photoexcited electrons}
\author{O.E. Raichev}
\author{F.T. Vasko}
\email{ftvasko@yahoo.com}
\affiliation{Institute of Semiconductor Physics, National Academy of Sciences of Ukraine,
Prospekt Nauki 45, Kiev, 03028, Ukraine}
\date{\today}

\begin{abstract}
Transient magnetotransport of two-dimensional electrons with
partially-inverted distribution excited by an ultrashort optical pulse
is studied theoretically. The time-dependent photoconductivity is
calculated for GaAs-based quantum wells by taking into account the
relaxation of electron distribution caused by non-elastic electron-phonon
interaction and the retardation of the response due to momentum relaxation
and due to a finite capacitance of the sample. We predict large-amplitude
transient oscillations of the current density and Hall field (Hall
oscillations) with frequencies corresponding to magnetoplasmon range,
which are initiated by the instability owing to the absolute negative
conductivity effect.

\end{abstract}

\pacs{73.50.Jt, 73.50.Pz, 73.50.Bk}

\maketitle

\section{Introduction}

The transient process under an abrupt turn on of the electric current
through a conducting sample is described by a simple exponential dependence
if the applied voltage is fixed, i.e. the load resistance is small and
the circuit effects are not essential. The characteristic temporal scale
of this process is determined by the momentum relaxation time, which
depends on the average energy of electrons and on the mechanisms of electron
scattering.$^{1,2}$ Such kind of exponential relaxation of electric current
in pure bulk Ge has been demonstrated more than 30 years ago.$^3$ To
investigate the transient processes, one can use ultrafast photoexcitation
of carriers instead of an abrupt switching of the applied voltage. Since
this excitation creates non-equilibrium electrons inside the conduction
band, the temporal dependence of the current contains a slow component
reflecting the energy relaxation of these electrons owing to quasielastic
scattering by acoustic phonons. This slow energy relaxation, corresponding
to the temporal scale which is much larger than the momentum relaxation time,
takes place for the electrons in the passive region (i.e. for the electrons
whose energies $\varepsilon$ are smaller than the optical phonon energy
$\hbar \omega_o$), because the interaction of these electrons with optical
phonons can be neglected at small temperatures $T \ll \hbar \omega_o$.
The slow temporal dependence of the transient photocurrent in this case
occurs for the same reason as the dependence of the photocurrent on
the excitation energy under stationary photoexcitation.$^4$

If the energies of excited electrons are close to $\hbar \omega_o$, see the
initial distribution B in Fig. 1 (a), the absolute negative conductivity (ANC)
should take place, because the non-equilibrium distribution $f_{\varepsilon}$ of
electrons in the passive region becomes inverted ($\partial f_{\varepsilon}/
\partial \varepsilon >0$) in a certain interval of energies near the upper boundary
of this region. Though such a possibility has been discussed a long time ago$^{5}$
for the regime of stationary photoexcitation in bulk samples, the ANC effect has
not been detected so far in this regime. The absence of the ANC under a stationary
photoexcitation is described by accumulation of low-energy electrons with time
(owing to slowness of interband recombination) so that the relative contribution
of the inverted high-energy part of electron distribution to the conductivity becomes
non-essential. In addition, the Coulomb scattering of high-energy electrons by
the low-energy ones leads to a rapid broadening of the initial narrow distribution
of photoexcited electrons in the energy space, thereby decreasing the contribution
of the inverted part of electron distribution. To date, the ANC has been realized
by means of electron heating by electric field in many-valley semiconductors$^6$
owing to intervalley redistribution of electrons, or when acting by microwave radiation
on two-dimensional (2D) electrons in a quantizing magnetic field.$^7$ Recently,
see Refs. 8, 9 and Ch. 11 in Ref. 10, it has been shown that the transient ANC,
which exists during a time interval of the order of momentum relaxation time,
can be achieved by ultrafast photoexcitation of electrons with energies close
to $\hbar \omega_o$. The theoretical description of this effect$^9$ has been
based on the kinetic theory taking into account temporal non-locality of the
response on the scale of momentum relaxation time and involving a detailed
consideration of inelastic scattering of high-energy electrons by acoustic
phonons. In this paper, we study the influence of classical magnetic fields
on the transient ANC in 2D samples with the geometry of a long Hall bar,
see Fig. 1 (b).

The main feature of the transient magnetotransport under consideration is the
appearance of temporal oscillations of the longitudinal conductivity and
transverse electric field (Hall field), whose frequencies are of the order of
the cyclotron frequency $\omega_c$ (Hall oscillations). Weak oscillations of
this kind should be always present in the transient response because of the
retardation of charge accumulation on the sides of the Hall bar. The
existence of the transient ANC leads to the instability which dramatically
modifies the transient oscillations. In the initial moments of time, when the
longitudinal current flows in the direction opposite to the applied field,
the Hall field increases in the direction opposite to its equilibrium one,
because the sign of the Lorentz force is changed in the ANC regime. For the
same reason, this increase is exponential: the charge accumulation on the sides
leads to further enhancement of this accumulation. In the subsequent moments
of time, when the partial inversion of the electron distribution is no longer
sufficient to provide the ANC, the system starts decharging, and, since the
system posesses a certain inertia, there appear large-amplitude oscillations
of the Hall field as well as of the longitudinal current (this current is
coupled to the Hall field). Both the current and the Hall field can change
their signs in the process of the oscillations. The damping of
such large-amplitude oscillations proceeds slower, which makes them persistent
on a nanosecond time scale.

\begin{figure}[ht]
\begin{center}
\includegraphics[scale=0.52]{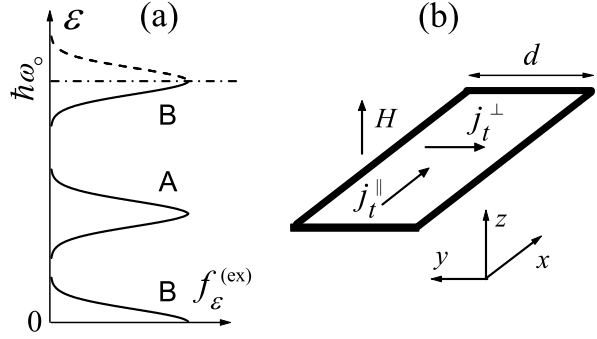}
\end{center}
\addvspace{-0.6 cm} \caption{($a$)
Initial electron energy distribution $f^{(ex)}_{\varepsilon}$ for the cases
of excitations away from the optical phonon energy (A) and near the optical
phonon energy (B). ($b$) Hall bar geometry and electric currents in the
presence of a magnetic field $H$ directed perpendicular to the 2D plane.}
\end{figure}

The paper is organized as follows. In Sec. II we derive general equations
for transient magnetotransport of electrons, which describe temporal dependence
of the longitudinal current and Hall field. Section III is devoted to a simple
model which makes it possible to solve these equations analytically and to describe
the main features of the transient response. In Sec. IV we present the results
of numerical calculations involving a detailed consideration of the evolution
of electron distribution. The discussion of the assumptions used and concluding
remarks are given in the last section. The Appendix A contains the expressions for
the transition probability and relaxation rate of 2D electrons interacting with
acoustic phonons. The Appendix B contains the formalism describing the retardation
of charge accumulation at the sides of the Hall bar.

\section{Temporal response}

We describe the response of photoexcited electrons to an electric
field ${\bf E}_t$ by representing the distribution function
$f_{\mathbf{p}t}$, which depends on the electron momentum $\mathbf{p}$
and time $t$, in the form $f_{\mathbf{p}t} = f_{\varepsilon t}+
\Delta f_{\mathbf{p}t}$, where $f_{\varepsilon t}$ is the
symmetric part of this function and $\Delta f_{\mathbf{p}t}$ is
the antisymmetric contribution induced by the field. Under the approximation
of weak electric field, when the heating of electrons by the field
is neglected, the symmetric part, which describes the energy distribution
of non-degenerate electrons, is governed by the kinetic equation
\begin{equation}
%1
\frac{\partial f_{\varepsilon t}}{\partial t}=J_{ac}\left( f_t \mid \varepsilon \right) +
J_{opt} \left( f_t \mid \varepsilon \right).
\end{equation}
The collision integral due to acoustic-phonon scattering, $J_{ac}$, can be
written as
\begin{equation}
%2
J_{ac}\left( f_t \mid \varepsilon \right) = \rho_{\ss 2D} \int_0^{\infty} d \varepsilon '
\left[ W_{\varepsilon' \varepsilon} f_{\varepsilon 't}- W_{\varepsilon
\varepsilon'} f_{\varepsilon t} \right],
\end{equation}
where $\rho_{\ss 2D}=m/\pi \hbar^2$ is the density of states of 2D electrons with
the effective mass $m$. The scattering probabilities $W_{\varepsilon' \varepsilon}$
and $W_{\varepsilon \varepsilon'}$ are determined by the deformation-potential (DA)
and piezoelectric (PA) interactions of electrons with acoustic phonons, see
Appendix A. These probabilities satisfy the requirement of detailed balance,
$W_{\varepsilon' \varepsilon}=W_{\varepsilon \varepsilon'}\exp [(\varepsilon' -
\varepsilon )/T]$, where $T$ is the phonon temperature. The collision integral
$J_{opt}$, which describes the interaction of non-degenerate electrons with
dispersionless optical phonons at $T \ll \hbar \omega_o$, can be represented
in a similar form:
\begin{eqnarray}
%3
J_{opt}\left( f_t \mid \varepsilon \right) = \nu_o \int d \varepsilon' \left[
\delta \left(\varepsilon -\varepsilon' +\hbar \omega_o \right) f_{\varepsilon' t}
\right. \nonumber \\
- \left. \delta \left(\varepsilon -\varepsilon' -\hbar \omega_o \right)
f_{\varepsilon  t} \right],
\end{eqnarray}
where $\nu_o$ is the rate of spontaneous emission of optical phonons by 2D
electrons (see, for example, Ref. 10).

Since the rate $\nu_o$ is typically
much larger than the rate of acoustic-phonon scattering, any electron
appearing in the active region ($\varepsilon > \hbar \omega_o$) after
photoexcitation or after acoustic-phonon absorption instantaneously
relaxes to a state in the passive region, with the energy $\varepsilon
-k \hbar \omega_o$, where $k$ is the number of emitted optical
phonons. In this approximation, the kinetic
equation (1) can be considered for the passive region only. To
carry out such a transformation, we first rewrite the term describing
departure of electrons from the state $\varepsilon$ in Eq. (2) as
$\rho_{\ss 2D} \int_0^{\hbar \omega_o} d \varepsilon ' \sum_{k=0}^{\infty}
W_{\varepsilon \varepsilon'+k \hbar \omega_o} f_{\varepsilon t}$.
Since the active region is empty owing to rapid emission of optical
phonons ($f_{\varepsilon t}=0$ at $\varepsilon > \hbar \omega_o$),
this term is considered in the region $\varepsilon < \hbar \omega_o$
only. For the same reason, the term corresponding to arrival of electrons
at the state $\varepsilon$ in Eq. (2) is written as $\rho_{\ss 2D}
\int_0^{\hbar \omega_o} d \varepsilon ' W_{\varepsilon' \varepsilon}
f_{\varepsilon' t}$. This term describes transitions of electrons
both to the states with $\varepsilon < \hbar \omega_o$ and to the
states with $\varepsilon > \hbar \omega_o$. As explained above,
in the latter case the electrons instantaneously relax to the states
with the energies $\varepsilon-k \hbar \omega_o$ in the passive region.
Therefore, in the presence of rapid spontaneous emission of optical
phonons the arrival term takes the form $\rho_{\ss 2D} \int_0^{\hbar \omega_o}
d \varepsilon ' \sum_{k=0}^{\infty} W_{\varepsilon' \varepsilon+
k \hbar \omega_o} f_{\varepsilon' t}$, where $\varepsilon < \hbar
\omega_o$. Finally, since $W_{\varepsilon \varepsilon'}$ becomes
exponentially small at $\varepsilon' - \varepsilon > \hbar \omega_o$,
see Appendix A, one should retain only the terms with $k=0$ and $k=1$ both
in the arrival and in the departure terms. As a result, Eq. (1) is reduced to
the following form:
\begin{eqnarray}
%4
\frac{\partial f_{\varepsilon t}}{\partial t}=\rho_{\ss 2D} \int_0^{\hbar \omega_o}
d \varepsilon ' \left[
(W_{\varepsilon' \varepsilon}+ W_{\varepsilon' \varepsilon+\hbar \omega_o})
f_{\varepsilon 't} \right. \nonumber \\
\left. - (W_{\varepsilon \varepsilon'}+ W_{\varepsilon \varepsilon'+\hbar \omega_o})
f_{\varepsilon t} \right].
\end{eqnarray}
The term with $W_{\varepsilon' \varepsilon+\hbar \omega_o}$ in this equation
becomes essential only if $\varepsilon'$ is close to $\hbar \omega_o$ and
$\varepsilon$ is close to zero. Similarly, the term with $W_{\varepsilon
\varepsilon'+\hbar \omega_o}$ becomes essential if $\varepsilon$ is
close to $\hbar \omega_o$ and $\varepsilon'$ is close to zero.

Equation (4) should be solved with the initial condition $f_{\varepsilon t=0}=
f_{\varepsilon}^{(ex)}$, where $f_{\varepsilon}^{(ex)}$ is determined
by the excitation pulse. If the initial ultrafast excitation creates
electrons with the distribution $F^{(ex)}_{\varepsilon}$,
we have $f^{(ex)}_{\varepsilon}=\theta(\hbar \omega_o-
\varepsilon) \sum_{k=0}^{\infty} F^{(ex)}_{\varepsilon + k \hbar \omega_o}$, where
the terms with $k \neq 0$ describe the electrons instantaneously transferred
to the passive region via optical phonon emission. Note that Eq. (4)
satisfies the density conservation requirement implying that
the electron density $\rho_{\ss 2D} \int_0^{\hbar \omega_o} d \varepsilon
f_{\varepsilon t}$ does not depend on time and equal to the excited density
$n_{ex}=\rho_{\ss 2D} \int_0^{\hbar \omega_o} d \varepsilon f^{(ex)}_{\varepsilon}$.

The weak antisymmetric part $\Delta f_{\mathbf{p}t}$, in the presence of a
stationary magnetic field $\bf H$ (${\bf E}\bot{\bf H}$), is governed by the
linearized kinetic equation
%5
\begin{eqnarray}
\left(\frac{\partial}{\partial t}+\frac{e}{c}[{\bf v}\times{\bf H}]
\! \cdot \! \frac{\partial}{\partial{\bf p}}\right)\Delta f_{{\bf p}t}+
e{\bf E}_t \! \cdot \! \frac{\partial f_{\varepsilon t}}{\partial {\bf p}}
\simeq \! -\nu _\varepsilon \Delta f_{\mathbf{p}t},
\end{eqnarray}
where $e$ is the electron charge, $c$ is the velocity of light, and ${\bf v}={\bf p}/m$
is the electron velocity. The momentum relaxation rate on the right-hand side of Eq. (5)
is the sum of the rate of quasielastic scattering of electrons by acoustic phonons,
$\nu_\varepsilon^{\ss (ac)}$, (see Appendix A) and the rate of spontaneous emission of
optical phonons, $\nu_o \theta(\varepsilon-\hbar \omega_o)$.
The exact solution of Eq. (5) is given by
\begin{eqnarray}
%6
\Delta f_{\mathbf{p}t}=e \int_0^t dt^{\prime } ~e^{-\nu_\varepsilon (t-t')} \mathbf{v}\cdot
{\bf K}_{tt'} \left( -\frac{\partial f_{\varepsilon t'}}{\partial \varepsilon }\right), \\
{\bf K}_{tt'} \equiv \mathbf{E}_{t'}\cos\omega_c(t-t^{\prime })
+ \frac{[ \bomega_c\times\mathbf{E}_{t'}]}{\omega _c}\sin\omega_c(t-t^{\prime }) ,
\nonumber
\end{eqnarray}
where $\bomega_c =|e|{\bf H}/mc$ is the cyclotron frequency vector.

The current density is given by the standard formula, ${\bf j}_t=(2/L^2)
\sum_{\mathbf{p}}\mathbf{v}\Delta f_{\mathbf{p}t}$, where
$L^2$ is the normalization square. Using Eq. (6) and performing the
averaging over the angle of ${\bf p}$, we write $\mathbf{j}_t$ as
%7
\begin{equation}
\mathbf{j}_t=\frac{e^2\rho_{\ss 2D}}{m}\int_0^t dt^{\prime } {\bf K}_{tt'}
\int_0^\infty d \varepsilon \varepsilon e^{-\nu
_\varepsilon (t-t^{\prime })}\left( -\frac{\partial f_{\varepsilon t^{\prime
}}}{\partial \varepsilon }\right).
\end{equation}
The linear response of electron system to the electric field ${\bf E}_{t}$
is described by the time-dependent conductivity tensor $\widehat\sigma_{tt'}$
introduced according to the non-local relation $\mathbf{j}_t=\int_0^t dt'
\widehat\sigma_{tt'}{\bf E}_{t'}$. The diagonal and non-diagonal components
of this tensor, $\sigma_{tt'}^{\ss \|}$ and $\sigma_{tt'}^{\ss \bot}$, are
\begin{eqnarray}
%8
\left|\begin{array}{c}\sigma_{tt'}^{\ss \|} \\ \sigma_{tt'}^{\ss\bot}\end{array}
\right| =\frac{e^2 \rho_{\ss 2D}}{m}\left| \begin{array}{c}
\cos \omega _c(t-t^{\prime }) \\ \sin \omega _c(t-t^{\prime })\end{array}
\right| \\
\times \int_0^{\hbar \omega _o} d \varepsilon \varepsilon
e^{-\nu _\varepsilon (t-t^{\prime })}\left( -\frac{\partial
f_{\varepsilon t^{\prime }}}{\partial \varepsilon }\right). \nonumber
\end{eqnarray}
The contribution of the active region is neglected in this equation,
because this region is depleted of electrons owing to rapid emission
of optical phonons. Accordingly, the scattering rate $\nu _\varepsilon$
standing in Eq. (8) is equal to the acoustic-phonon scattering rate
$\nu_\varepsilon^{\ss (ac)}$ calculated in Appendix A.

Below we consider a sample of Hall bar geometry, a 2D strip of width $d$
with the in-plane current density ${\bf j}_t=(j_t^{\ss\|},~j_t^{\ss\bot})$,
where $\|$ and $\bot$ components of ${\bf j}_t$ are referred to the
coordinate system associated with the geometry of the Hall bar
(Fig. 1 b). The transverse current density $j_t^{\ss \bot}$ is not equal
to zero and describes the transient process of charge accumulation on the
sides (edges) of the Hall bar. Owing to near-edge localization of
the magnetoinduced charges and current continuity, one can use the
homogeneous current vector. Under the assumption of high resistance
of the photoexcited electron gas in comparison to the load resistance
of the circuit, we have $\mathbf{E}_t =(E^{\ss\|}, E_t^{\ss\bot})$,
where the longitudinal field $E^{\ss\|}$ is determined by the applied
voltage and remains time-independent. The Hall field $E_t^{\ss\bot}$
depends on time because of the charge accumulation process mentioned
above. The components of the current density vector are written
through the components of conductivity tensor (8) as follows:
%9,10
\begin{eqnarray}
j_t^{\ss\bot}=\int_0^tdt'\left( \sigma_{tt'}^{\ss\|} E^{\ss\perp}_{t'}+
\sigma_{tt'}^{\ss \perp} E^{\ss\Vert }\right) , \\
j_t^{\ss\|}=\int_0^tdt'\left( \sigma _{tt^{\prime}}^{\ss\|}E^{\ss\|} -
\sigma _{tt'}^{\ss\perp }E^{\ss\perp }_{t'}\right) \equiv \sigma_t^{\ss eff}
E^{\ss\Vert }  ,
\end{eqnarray}
where we have introduced the effective conductivity $\sigma_t^{\ss eff}$.
To obtain a closed equation for the Hall field, one should describe
the charge accumulation at the sides of the Hall bar. This leads to
the approximate equation
%11
\begin{equation}
j_t^{\ss\bot} \simeq -C_{\ss \bot} \frac{dE_t^{\ss\bot}}{dt},
\end{equation}
where $C_{\ss \bot}=\alpha \epsilon d$ is the effective capacitance
proportional to the dielectric permittivity $\epsilon$, and $\alpha$
is a small numerical factor. The derivation of Eq. (11) and the estimate
of $\alpha$ are given in Appendix B.

In summary, to describe the linear response of the system, one has first to
solve Eq. (4) and determine the energy distribution $f_{\varepsilon t}$.
Then, $\sigma_{tt'}^{\ss \|}$ and $\sigma_{tt'}^{\ss \bot}$ are calculated
according to Eq. (8). Using them in Eq. (9) and applying Eq. (11), one finds
the Hall field $E^{\ss\perp}_{t}$, which is proportional to the time-independent
longitudinal field $E^{\ss\|}$. Finally, the longitudinal current is expressed
through the effective conductivity from Eq. (10): $\sigma_t^{\ss eff}=
\int_0^tdt'\left( \sigma_{t t^{\prime}}^{\ss\|} - \sigma _{tt'}^{\ss\perp}
E^{\ss\perp}_{t'}/E^{\ss\|} \right)$.

\section{Analytical Approach}

Before presenting the results of numerical solution of Eq. (9), we
discuss an approximation which allows one to understand essential
features of the time-dependent response by means of analytical
consideration. First of all, we neglect the energy dependence of
the momentum relaxation time, replacing $\nu_{\varepsilon}$
in Eq. (8) by a constant $\nu = \nu_{\hbar \omega_o}$. Note that the
calculated energy dependence $\nu_{\varepsilon}$ in the interval
$0 < \varepsilon < \hbar \omega_o$ is not strong, except for the
low-energy region (see Fig. 2). The integral over energy in Eq. (8)
in this case is taken by parts, with the result

\begin{figure}[ht]
\begin{center}
\includegraphics[scale=0.48]{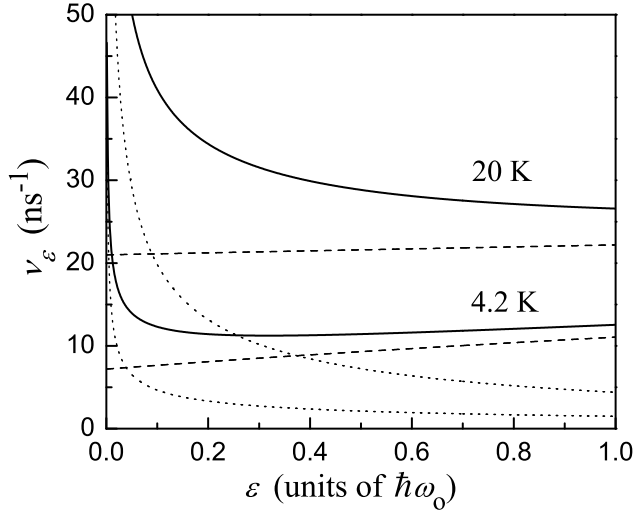}
\end{center}
\addvspace{-0.8 cm}\caption{
Energy dependence of momentum relaxation rates for GaAs quantum well
of width 10 nm for $T=4.2$ K and $T=20$ K. The dashed and dotted lines show
the partial contributions of DA and PA scattering mechanisms, respectively.}
\end{figure}

\begin{eqnarray}
%12
\left|\begin{array}{c}\sigma_{tt'}^{\ss \|} \\ \sigma_{tt'}^{\ss\bot}\end{array}
\right| =\frac{e^2 n_{ex}}{m}\left| \begin{array}{c}
\cos \omega _c(t-t^{\prime })  \\ \sin \omega _c(t-t^{\prime })\end{array}
\right| e^{-\nu (t-t^{\prime })} g_{t'}~, \\
g_t=\left( 1- \frac{\rho_{\ss 2D}
\hbar\omega_o}{n_{ex}}f_{\hbar\omega_o t}\right).~~~~~~~~~~~ \nonumber
\end{eqnarray}
The evolution of the energy
distribution enters the conductivity tensor (12) through the dimensionless
function $g_t$, which depends on the distribution function at the
boundary of the passive region. After substituting the expression (11) to
Eq. (9) with $\sigma_{tt'}^{\ss \|}$ and $\sigma_{tt'}^{\ss\bot}$ given by
Eq. (12), it is convenient to differentiate the equation obtained over $t$
twice. As a result, Eq. (9) is reduced to the differential equation
\begin{eqnarray}
%13
\frac{d^3 E^{\ss\perp}_{t}}{d t^3} + 2 \nu \frac{d^2 E^{\ss\perp}_{t}}{d t^2}
+(\omega_c^2+ \nu^2 + \Omega^2 g_t) \frac{d E^{\ss\perp}_{t}}{d t} \nonumber \\
+ \Omega^2 \left(\nu g_t + \frac{d g_{t}}{d t} \right) E^{\ss\perp}_{t}
+ \Omega^2 \omega_c g_t E^{\ss\|}=0
\end{eqnarray}
with the initial conditions $E^{\ss\perp}_{t}= d E^{\ss\perp}_{t}/d t =
d^2 E^{\ss\perp}_{t}/d t^2 =0$ at $t=0$. We have introduced a characteristic
frequency
\begin{equation}
%14
\Omega=\sqrt{\frac{e^2 n_{ex}}{m C_{\ss \bot}}}
\end{equation}
determined by the capacitance and electron density. The longitudinal current,
given in the integral form by Eq. (10), can be expressed through the derivatives
of $E^{\ss\perp}_{t}$ with the use of Eqs. (9), (10), (11), and (12):
\begin{eqnarray}
%15
j_t^{\ss\|}=-\frac{e^2 n_{ex}}{m \Omega^2 \omega_c} \left( \frac{d^2 E^{\ss
\perp}_{t}}{d t^2} + \nu \frac{d E^{\ss\perp}_{t}}{d t}
+ \Omega^2 g_t E^{\ss\perp}_{t} \right).
\end{eqnarray}

If the electrons are excited considerably below the optical phonon energy, see the
initial distribution A in Fig. 1 (a), one has $g_t=1$. In this case
Eq. (13) is solved analytically. The solution shows a three-mode behavior
according to
\begin{eqnarray}
%16
E^{\ss\perp}_{t} = - E^{\ss\|} \frac{\omega_c}{\nu}
\left(1 - c_1 e^{s_1 t} - c_2 e^{s_2 t} - c_3 e^{s_3 t} \right),~~~~ \\
c_1=\frac{s_2 s_3}{(s_1-s_2)(s_1-s_3)}~,~~
c_2=\frac{s_1 s_3}{(s_2-s_1)(s_2-s_3)}~, \nonumber \\
c_3=\frac{s_1 s_2}{(s_3-s_1)(s_3-s_2)}~,~~~~~~~~~~~~~~~ \nonumber
\end{eqnarray}
where $s_{1-3}$ are the roots of the characteristic equation
$s^3+ 2\nu s^2+( \omega_c^2+ \nu^2 + \Omega^2) s +\nu \Omega^2=0$.
Under the approximation $\nu^2 \ll \omega_c^2 + \Omega^2$, the
solution (16) is rewritten as
\begin{eqnarray}
%17
E^{\ss\perp}_{t} \simeq  - E^{\ss\|} \frac{\omega_c}{\nu} \left[ 1 -
\exp \left(-\frac{\Omega^2}{\Omega_c^2} \nu t \right) \right. \\
\left. -\frac{\Omega^2 \nu}{\Omega_c^3} \exp \left(-\frac{\omega_c^2+
\Omega^2/2}{\Omega_c^2} \nu t \right) \sin (\Omega_c t) \right], \nonumber
\end{eqnarray}
where $\Omega_c=\sqrt{\omega_c^2 + \Omega^2}$. The expression (17) describes
the increase of the Hall field from 0 to its equilibrium value $- E^{\ss\|}
\omega_c/\nu$ with a characteristic time $(\Omega_c/\Omega)^2 \nu^{-1}$ and weak
oscillations of this field with the frequency $\Omega_c$. The oscillations are
exponentially damped on the time scale of $\nu^{-1}$, though the damping is
suppressed at $(\omega_c/\Omega)^2 \ll 1$. In the case of $\omega_c
\gg \Omega$, which still can be realized in the classical region of magnetic
fields if the excitation density is low enough, the increase of the Hall
field appears to be much slower than the damping of the oscillations. The
longitudinal current shows a similar evolution, which is obvious from the
relation (15).

If the electrons are excited close to the optical phonon energy (see
the distribution B in Fig. 1 (a)), the function $g_t$ is not equal
to unity and can be negative at the initial moments of time, owing to
the partial inversion of the electron distribution. As the excited
electrons relax and go away from the boundary of the passive region,
$g_t$ changes its sign from negative to positive at some instant $t=t_0$
and approaches 1 at $t \rightarrow \infty$. Although Eq. (13) cannot
be solved analytically in the general case, the basic features of the
response can be determined by using the model step-like function
\begin{equation}
%18
g_t= \left\{ \begin{array}{c} g_0,~~t<t_0 \\ 1, ~~ t > t_0  \end{array} \right. ,
\end{equation}
where $g_0$ is a negative constant. Substituting the expression (18) into
Eq. (13), one can find that at $t > t_0$ the solution (16) is valid again.
However, the coefficients $c_{1-3}$ should be found by means
of matching this solution to the solution at $t < t_0$, which has the form
\begin{eqnarray}
%19
\left. E^{\ss\perp}_{t} \right|_{t < t_0} = - E^{\ss\|} \frac{\omega_c}{\nu}
\left(1 - d_1 e^{p_1 t} - d_2 e^{p_2 t} - d_3 e^{p_3 t} \right),
\end{eqnarray}
and $p_{1-3}$ are the roots of the equation
$p^3+ 2\nu p^2+( \omega_c^2+ \nu^2 + \Omega^2 g_0) p +\nu \Omega^2 g_0=0$.
The coefficients $d_{1-3}$ are expressed through $p_{1-3}$ in the same way
as the coefficients $c_{1-3}$ are expressed through $s_{1-3}$, see Eq. (16).
The rules of the matching are derived from integration of Eq. (13) across the
point $t=t_0$ and imply continuity of the Hall field $E^{\ss\perp}_{t}$ and
its first time derivative, while the second derivative has a finite step,
$\left. d^2 E^{\ss\perp}_{t}/d t^2 \right|_{t=t_0-0}^{t=t_0+0}=-(1-g_0)E^{\ss
\perp}_{t=t_0}$, which provides the continuity of the current given
by Eq. (15). It is essential that at least one of the rates $p_{1-3}$
has a positive real part, which describes exponential increase of the
Hall field in the interval $t < t_0$. This is a manifestation of the
instability generated by the ANC effect. It is also important that the
sign of the increasing Hall field is opposite to its equilibrium sign
because of inversion of the direction of current in the ANC interval.
The strong enhancement of the Hall field at $t < t_0$ initiates
large-amplitude oscillations of this field (Hall oscillations) in the
region $t > t_0$. The oscillations of the longitudinal current are also
dramatically enhanced.

\begin{figure}[ht]
\begin{center}
\includegraphics[scale=0.46]{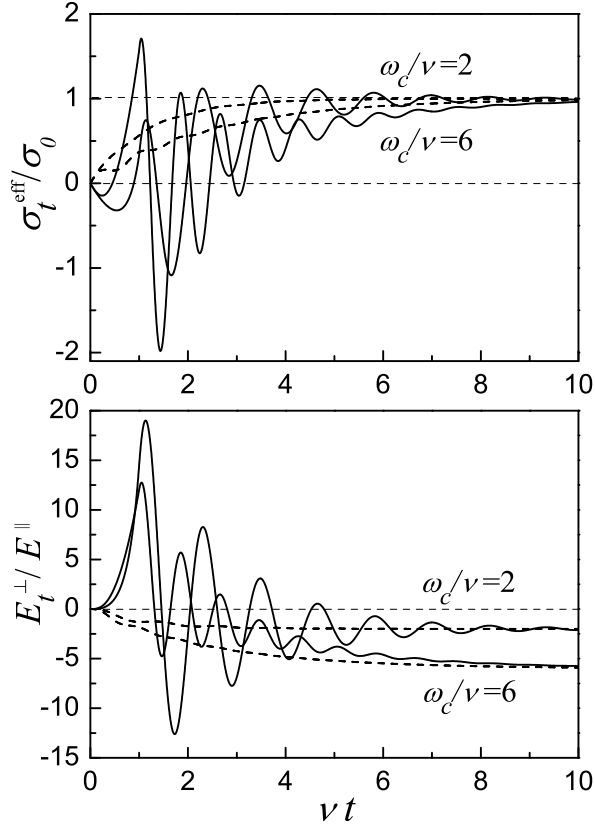}
\end{center}
\addvspace{-0.8 cm} \caption{Evolution of the effective conductivity
$\sigma_t^{\ss eff}$ and Hall field $E^{\ss\perp}_{t}$ calculated within
the approximation described by Eqs. (12) and (18) at $\Omega/\nu=5$, $t_0=
\nu^{-1}$, and $g_0=-1$ for two values of cyclotron frequency, $\omega_c/\nu=2$
and $\omega_c/\nu=6$. The dashed lines show the corresponding evolution for
the case $g_t=1$, when the excited electrons appear considerably below
the optical phonon energy.}
\end{figure}

Figure 3 demonstrates some examples of temporal dependence of $E^{\ss\perp}_{t}$
and $\sigma_t^{\ss eff}$ (expressed in units of $\sigma_0=e^2 n_{ex}/m \nu$)
calculated within the analytival approach described here. We have chosen
$\Omega/\nu=5$, $t_0=\nu^{-1}$ and $g_0=-1$. For comparison, we also plot
the corresponding temporal dependence at $g_t=1$, when the electrons are
excited below the optical phonon energy. The oscillations in this case
also exist, but they are weak and superimposed on a smooth exponential
relaxation dependence. On the other hand, the oscillations generated
by the ANC instability are strong and remain visible even at $\nu t \simeq 10$.
Estimating $\nu \simeq 12$ ns$^{-1}$, see Fig. 2 for 4.2 K, one can
conclude that these oscillations persist over a nanosecond interval of
time after the excitation. The frequency of the oscillations increases with
the increase of the magnetic field and is estimated as $\Omega_c=\sqrt{\omega_c^2
+ \Omega^2}$. According to Eq. (14) and to the estimate of $C_{\ss \bot}$, the
frequency $\Omega$ is of the order of 2D plasmon frequency $\omega_q$ at
wavenumbers $q$ corresponding to the
inverse width of the Hall bar, $q \sim 1/d$. For this reason, $\Omega_c$ is
identified with a 2D magnetoplasmon frequency. The appearance of the
large-amplitude Hall oscillations can therefore be considered as excitation
of 2D magnetoplasmons owing to the ANC instability. The amplitude of the
oscillations exponentially increases with the increase of the frequency
$\Omega$ and with the increase of the absolute value of $g_0$. We also
note that the parameters used in the calculations are physically
reasonable. Indeed, if $\nu \simeq 12$ ns$^{-1}$ (Fig. 2), then $\Omega/\nu=5$
corresponds, for example, to $C_{\ss \bot} \simeq 0.1$ cm and $n_{ex}
\simeq 10^{11}$ cm$^{-2}$ (or $C_{\ss \bot} \simeq 0.01$ cm and $n_{ex}
\simeq 10^{10}$ cm$^{-2}$) for GaAs wells, while the ratios $\omega_c/\nu=2$
and $\omega_c/\nu=6$ approximately correspond to the classical magnetic
fields of 0.01 T and 0.03 T, respectively.

\section{Numerical Results}

The model consideration given above ignores a detailed evolution of
electron distribution after the photoexcitation. Below we present the
results of a careful consideration based on the numerical solution of Eq. (9)
with $\sigma_{tt'}^{\ss \|}$ and $\sigma_{tt'}^{\ss \bot}$ given by Eq. (8).
We have used standard material parameters of GaAs, which can be found,
for example, in Refs. 1 and 11 (see also our paper, Ref. 9). To find the electron
distribution $f_{\varepsilon t}$ from Eq. (4), we assume that the optical pulse
creates electrons with a Gaussian energy distribution $F^{(ex)}_{\varepsilon} \propto
\exp [-(\varepsilon-\varepsilon_{ex})^2/\Delta^2]$ of a characteristic half-width
$\Delta$, centered at the excitation energy $\varepsilon_{ex}$. Assuming,
for example, that $\varepsilon_{ex}=\hbar \omega_o$ and $(\Delta/\hbar
\omega_o)^2 \ll 1$, we obtain the initial electron energy distribution
in the passive region in the form of two half-peaks also shown in Fig. 1 (a):
\begin{equation}
%20
f^{(ex)}_{\varepsilon} \propto \exp [-(\varepsilon-\hbar \omega_o)^2/
\Delta^2]+ \exp [-\varepsilon^2/\Delta^2],
\end{equation}
see the discussion after Eq. (4). The numerical solution of Eq. (4) has been
carried out by iterations in the time domain. The evolution of the electron
distribution, calculated for 10 nm wide GaAs quantum well at 4.2 K, is shown
in Fig. 4. This evolution is similar to that calculated for bulk samples in
Ref. 9.

\begin{figure}[ht]
\begin{center}
\includegraphics[scale=0.46]{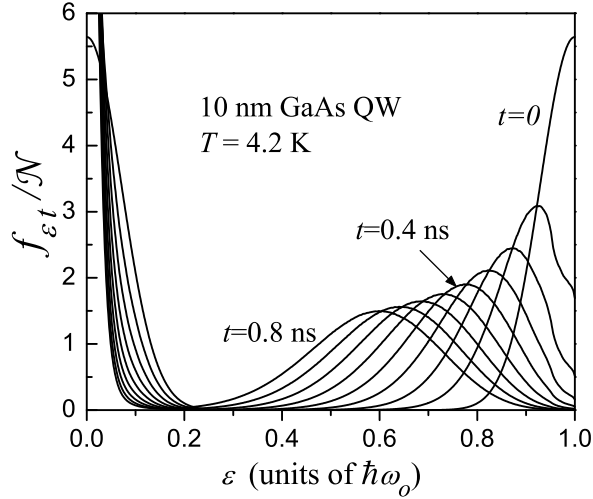}
\end{center}
\addvspace{-0.6 cm}
\caption{Temporal evolution of electron energy distribution for the case
of initial distribution (20) with $\Delta = 0.1$ $\hbar \omega_o$, calculated
for 10 nm wide GaAs quantum well at 4.2 K. The distribution functions are
plotted with the time interval of 0.1 ns and normalized by ${\cal N}=
n_{ex} [\rho_{\ss 2D} \hbar\omega_o]^{-1}$.}
\end{figure}

Figure 5 shows the temporal dependence of the effective conductivity
$\sigma_t^{\ss eff}$ and Hall field $E^{\ss\perp}_{t}$ for 10 nm wide
GaAs quantum well at 4.2 K, calculated for the same parameters of the
excitation. The effective conductivity is expressed in units of
$\sigma_0$ defined in Sec. III. The temporal dependence of
$\sigma_t^{\ss eff}$ at zero magnetic field is shown by a dashed line,
and is similar to that calculated for bulk samples in Ref. 9.
The evolution of $\sigma_t^{\ss eff}$ and $E^{\ss\perp}_{t}$ appears
to be very sensitive to the cyclotron frequency and characteristic
frequency $\Omega$ [see Eq. (14)] because of the initial exponential
increase of the current and Hall field. We have chosen these frequencies
in such a way that the absolute values of $\sigma_t^{\ss eff}$ and
$E^{\ss\perp}_{t}$ in Fig. 5 are not too large. By varying the
parameters, one can obtain a very strong (several orders of magnitude)
enhancement of $\sigma_t^{\ss eff}$ and $E^{\ss\perp}_{t}$, but the
qualitative picture of the damped oscillations remains the same.
The increase of the magnetic field leads to increasing frequency of
the oscillations, while the amplitude of the oscillations decreases and
the relaxation slows down. The increase of the excitation density, which
leads to increasing frequency $\Omega$, exponentially increases the
amplitude of the oscillations.$^{12}$ This influence of the parameters on
the evolution is also described by the simple model investigated in the
previous section.

\begin{figure}[ht]
\begin{center}
\includegraphics[scale=0.46]{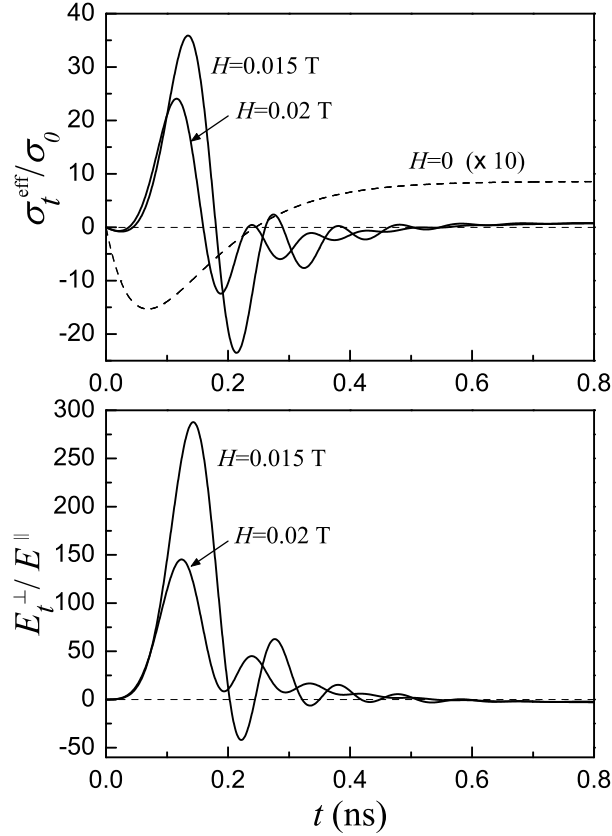}
\end{center}
\addvspace{-0.8 cm}\caption{Evolution of the effective conductivity
$\sigma_t^{\ss eff}$ and Hall field $E^{\ss\perp}_{t}$ at $\Omega=50$
ns$^{-1}$ in the magnetic fields $H=0.015$ T and 0.02 T, for the case
of initial distribution (20) with $\Delta = 0.1$ $\hbar \omega_o$. The
dashed line shows $\sigma_t^{\ss eff}$ (multiplied by 10) for
zero magnetic field.}
\end{figure}

The general features of the evolution are not modified if the electrons
are excited by a shorter optical pulse, which results in energy broadening
of the initial distribution. Figure 6 shows the temporal dependence of
$\sigma_t^{\ss eff}$ and $E^{\ss\perp}_{t}$ calculated for the case of
initial distribution (20) with $\Delta = 0.2$ $\hbar \omega_o$.
A comparison of this figure to Fig. 5 also demonstrates the increase
of the oscillation frequency and a suppression of the damping as a
result of the increased $\Omega$ (see the discussion of Eq. (17)).

\begin{figure}[ht]
\begin{center}
\includegraphics[scale=0.46]{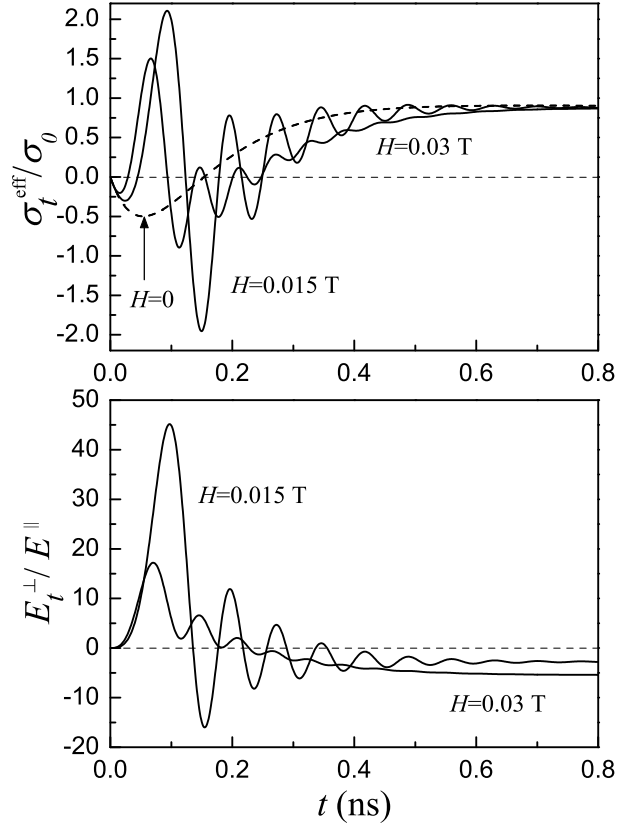}
\end{center}
\addvspace{-0.8 cm} \caption{Evolution of the effective conductivity
$\sigma_t^{\ss eff}$ and Hall field $E^{\ss\perp}_{t}$ at $\Omega=80$
ns$^{-1}$ in the magnetic fields $H=0.015$ T and 0.03 T, for the case
of initial distribution (20) with $\Delta = 0.2$ $\hbar \omega_o$.
The dashed line shows $\sigma_t^{\ss eff}$ for zero magnetic field.}
\end{figure}

\section{Conclusions}

We have described the transient classical magnetotransport of electrons in a
long Hall bar after ultrafast interband photoexcitation, and calculated the
temporal dependence of the current and Hall field. Investigating the modification
of the response due to accumulation of the edge charges forming the transverse
(Hall) voltage, we have found the oscillations of both the current and the Hall
field. The amplitude and duration of the oscillations are dramatically enhanced
if the energies of the excited electrons are in the vicinity of the optical
phonon energy. This is the case when the oscillations are triggered by the
instability associated with the partial inversion of electron distribution
(the ANC effect). Although the numerical calculations have been carried out
here for non-doped GaAs quantum wells, similar effects should be expected for
non-doped bulk samples, because the qualitative features of the energy relaxation
and non-local temporal response of electron system do not depend on dimensionality.

Now we discuss the assumptions used. The main approximation is the
consideration of electron scattering by phonon modes only. The elastic
scattering by inhomogeneities can be neglected in the case of non-doped
quantum wells with high-quality interfaces. In any case, it is not
difficult to include this scattering into consideration, because it
does not contribute to the energy relaxation (see Eq. (4)), and can
lead only to an increase in the momentum relaxation rate standing
in Eq. (8). However, the damping of the oscillations in this case becomes
stronger, and it is always better to avoid this additional scattering
by using clean samples. A more important restriction is the neglect of
electron-electron interaction, which should dominate the energy relaxation
of photoexcited 2D electrons at the densities $n_{ex} > 10^{10}$ cm$^{-2}$.
Since the electron-electron interaction leads to a faster relaxation, it is
expected to shorten the time interval where the exponential increase of the
current and Hall field takes place. However, this interaction cannot cancel
the large-amplitude oscillations of the current and Hall field. In this
connection, we note that these oscillations have been examined in
Sec. III on the basis of a model that is not sensitive to the detailed
description of the energy relaxation. The only essential point is the
existence of partial inversion of electron distribution during a finite
interval of time after the photoexcitation, sufficient for realization of
the ANC regime. For this reason, the position and energy broadening
of the initial electron distribution, which are determined by the
parameters of photoexcitation pulse, appear to be more important than
the actual energy relaxation mechanisms.

Let us discuss the other approximations. The general formalism has been
based on the classical Boltzmann equation. This is sufficient for the
subject of our study, because the intervals of times under consideration
considerably exceed the quantum broadening times $\hbar/\varepsilon$, and
the magnetic field is weak enough to neglect the Landau quantization. We
have ignored the existence of holes created in the valence band by the
optical pulse. This is possible because of smallness of the hole mobility,
so the contribution of the holes to the transport can be neglected.
To describe the momentum relaxation by electron-phonon scattering, we
have used the elastic approximation. In quantum wells, this is suitable
for description of electrons whose energies $\varepsilon$ are much larger than
the characteristic energy $\hbar s \pi/a$ associated with phonon momentum
normal to the layer (here $s$ is the sound velocity and $a$ is the well width).
For typical parameters $s \simeq 5 \times 10^5$ cm/s and $a \simeq 10$ nm, the
energy $\hbar s \pi/a$ is around 1 meV. Therefore, the assumed condition is
satisfied in our calculations for the photogenerated high-energy electrons
which give the main contribution to the conductivity. Next, since we have
neglected the transverse inhomogeneity of the currents and fields, the
equations (9)-(11) are not applicable for description of electrons in the
vicinity of the Hall bar edges. Nevertheless, these equations are valid in
the main part of the Hall bar, where the inhomogeneous corrections to the
currents and fields are small (see Appendix B).

We also note that in the ANC regime one should consider a possibility
for the development of a spatial instability$^{13}$ both along and across
the Hall bar. Uncovering the conditions for existence and properties of such
an instability requires a special investigation based on the formalism of
non-homogeneous kinetic equation. The spatially-inhomogeneous electron
distribution owing to the ANC effect is essential in the stationary regime,
and it is realized for 2D electrons under microwave excitation in the
quantizing magnetic field.$^7$ Nevertheless, since we consider the transient
response, one may expect that the results of this paper will remain valid
for the samples of small size, where the spatially-inhomogeneous distribution
is not developed during the short interval of time corresponding to the
exponential increase of the current and Hall field.

Finally, we would like to point out that the Hall oscillations studied in
this paper can be viewed as 2D magnetoplasmons with small wavenumbers
determined by the width of the Hall bar. To detect them in experiment, it
is necessary to have sub-nanosecond temporal resolution, which is attainable
for standard all-electrical measurements. Owing to the strong amplification
of the oscillations by the ANC instability, the excitation technique based
on the ultrafast optical pump can be applied along with the conventional
techniques$^{14}$ of magnetoplasmon excitation in 2D layers.

\appendix

\section{Relaxation rates}

Below we present the expressions for $\nu _\varepsilon ^{\ss (ac)}$ and
$W_{\varepsilon \varepsilon'}$ determining the momentum and energy
relaxation of 2D electrons under acoustic-phonon scattering. Both these
quantities can be written as the sums of partial contributions caused
by the DA and PA mechanisms of interaction. The relaxation rate
$\nu_{\varepsilon}^{\ss (ac)}$ is calculated in the elastic approximation:
\begin{eqnarray}
%A1
\nu _{\varepsilon} ^{\ss (ac)}=\frac{2 \pi }{\hbar} %\sum_{\mathbf{p'}, \q_{\bot}}
\int \frac{d {\bf p}'}{(2 \pi \hbar)^2} \int_{-\infty}^{\infty}
\frac{d q_{\ss \bot}}{2 \pi} M(q_{\ss \bot}) ( 1- \cos \varphi )
\nonumber \\
\times \delta( \varepsilon_{p} -\varepsilon _{p'})
\sum_{i=\ss DA,PA} \sum_{\lambda=l,t}
\overline{ |C_{\lambda {\bf Q}}^{\ss (i)}|^2 } (2N_{\lambda Q}+1),~~
\end{eqnarray}
where $\varepsilon=\varepsilon_{p}=p^2/2m$, ${\bf Q}=({\bf q},q_{\ss \bot})$ is the
phonon wave vector, ${\bf q}= ({\bf p}-{\bf p}')/\hbar$ is the in-plane component
of this vector, $\varphi$ is the angle between ${\bf p}$ and ${\bf p}'$,
$N_{\lambda Q}$ are the Planck occupation numbers of the longitudinal
($\lambda=l$) and transverse ($\lambda=t$) acoustic phonon modes, and
$\overline{ |C_{\lambda {\bf Q}}^{\ss (i)}|^2 }$ are the matrix
elements of interaction, averaged over the angle of electron momentum
${\bf p}$ in the 2D plane. This averaging is essential for the PA interaction,
which is sensitive to orientation of the vector ${\bf Q}$ with respect to
crystallographic axes.$^{15}$ In the explicit form,
\begin{equation}
%A2
\overline{ |C_{\lambda {\bf Q}}^{\ss (DA)}|^2 } =
\frac{ \hbar {\cal D}^2 Q}{2 \rho s_\lambda } \delta_{\lambda, l}
\end{equation}
and
\begin{equation}
%A3
\overline{ |C_{\lambda {\bf Q}}^{\ss (PA)}|^2 } =
\frac{ \hbar (e h_{14})^2}{2 \rho s_\lambda Q} A_{\lambda {\bf Q}}~,
\end{equation}
where $s_\lambda$ are the sound velocities, $\rho$ is the material density,
${\cal D}$ is the deformation-potential constant, and $h_{14}$ is the
piezoelectric coefficient. The polarization factors $A_{\lambda {\bf Q}}$ for
(100)-grown 2D layers are$^{16}$
\begin{equation}
%A4
A_{l {\bf Q}}= \frac{9}{2} \frac{q^2_{\ss \bot} q^4}{Q^6}~,~~~
A_{t {\bf Q}}= 4 \frac{q^4_{\ss \bot} q^2}{Q^6} + \frac{1}{2} \frac{q^6}{Q^6}.
\end{equation}
Finally, $M(q_{\ss \bot})=| \langle 0| e^{iq_{\bot}z}|0 \rangle |^2$
is the squared matrix element of a plane-wave factor. This matrix element
characterizes the interaction of 2D electrons with 3D phonons and depends
on the confinement potential determining the ground state of 2D electrons,
$|0\rangle$. We apply the model of a deep square well of width $a$ leading
to the expression$^2$ $M(q_{\ss \bot })=[\sin(x)/x]^2/[1-(x/\pi)^2]^2$,
where $x=q_{\ss \bot} a/2$.

The introduction of the transition probability $W_{\varepsilon \varepsilon'}$
entering Eqs. (2) and (4) implies averaging of the kinetic equation over the
angle of momentum ${\bf p}$, according to $\partial f_{\varepsilon t}/\partial t
= (2 \pi)^{-1} \int_0^{2 \pi} d \varphi_{\bf p} \sum_{\bf p'} (W_{\bf p' p}
f_{\varepsilon' t}- W_{\bf p p'} f_{\varepsilon t})$, where $W_{\bf p p'}$ is
the probability of electron scattering from the state with momentum ${\bf p}$
to the state with momentum ${\bf p}'$. As a result,
\begin{eqnarray}
%A5
W_{\varepsilon \varepsilon'}= \frac{1}{\hbar^2}
\frac{{\rm sgn}(\varepsilon -\varepsilon')}{1-\exp[(\varepsilon'-\varepsilon)/T]}
\nonumber \\
\times \sum_{i=DA,PA} \sum_{\lambda=l,t}
\int_0^{\varphi_\lambda} \frac{d \varphi}{\pi}
M(q_{\ss \bot}) \overline{ |C_{\lambda {\bf Q}}^{\ss (i)}|^2 }
\frac{Q}{s_\lambda q_{\ss \bot} },
\end{eqnarray}
where $\overline{ |C_{\lambda {\bf Q}}^{\ss (i)}|^2 }$ and $M(q_{\ss \bot})$
are already defined above. However, the wave numbers $Q$, $q$, and
$q_{\ss \bot}$ standing in the corresponding equations should be now
considered as functions of energies, phonon polarization, and scattering
angle $\varphi$, according to $Q=|\varepsilon -\varepsilon'|/\hbar s_\lambda$,
$q= \sqrt{2m(\varepsilon +\varepsilon' - 2 \sqrt{\varepsilon \varepsilon'}
\cos\varphi)}/\hbar$, and $q_{\ss \bot}= \sqrt{Q^2-q^2}$. The integral
over the scattering angle in Eq. (A5) must be calculated numerically. The
upper limit of this integration, $\varphi_\lambda$, is determined from
the requirement $Q^2 > q^2$, which means $\varphi_\lambda = \pi$ at
$(\sqrt{\varepsilon}-\sqrt{\varepsilon'})^2 > 2ms^2_{\lambda}$ and $\varphi_\lambda=
\arccos \left( [\varepsilon+ \varepsilon'-(\varepsilon-\varepsilon')^2/2m
s^2_{\lambda}]/ 2\sqrt{\varepsilon \varepsilon'} \right)$ at $(\sqrt{\varepsilon}
-\sqrt{\varepsilon'})^2 < 2ms^2_{\lambda}$. In the case of very small energies,
$(\sqrt{\varepsilon}+ \sqrt{\varepsilon'})^2 < 2 m s^2_{\lambda}$, the limit
$\varphi_\lambda$ should be set at zero and the integral gives no contribution.

\section{Edge charge accumulation}

This appendix contains a simple formalism describing 
inhomogeneous distributions of charges, currents, and fields in a 
long Hall bar for a time-dependent problem. The homogeneous approximation, 
which is valid in the central part of the bar and leads to Eq. (11) for 
the transverse current describing edge charge accumulation, follows 
from the general consideration.  

Consider a semiconductor structure containing a 2D Hall bar of width $d$ and
length $L$ ($-d/2<y<d/2$ and $-L/2< x <L/2$) placed at a distance $h$ below
the surface of the medium with dielectric permittivity $\epsilon$.
If the retardation effects are neglected, the electrostatic
potential $\Phi_t(y,z)$ created by the 2D carriers satisfies the Poisson
equation
\begin{equation}
%B1
\left[\frac{\partial}{\partial z} \epsilon(z) \frac{\partial}{\partial z} +
\epsilon(z) \frac{\partial^2}{\partial y^2} \right] \Phi_t(y,z) =- 4 \pi e
\delta n_t(y) \delta(z+h),
\end{equation}
where $\epsilon(z)=\epsilon$ at $z<0$, $\epsilon(z)=1$ at $z>0$,
$\delta n_t(y)=n_t(y)-p$ is the excess density of free electrons, and $p$
is the density of holes which is assumed to be homogeneous and time-independent
(because of low mobility of holes, this is possible in the short time interval
under consideration). The Poisson equation is two-dimensional because the
bar is long, $d \ll L$, and the electron density depends only on the
transverse coordinate $y$. If $\epsilon \gg 1$, one can use the Newmann's
boundary condition $\left. \partial \Phi_t(y,z)/\partial z \right|_{z=-0}=0$
at the surface and rewrite Eq. (1) in the integral form
\begin{equation}
%B2
\Phi_t(y,z) =- \frac{e}{\epsilon} \int_{-d/2}^{d/2} d y'
\left[ \ln[(y-y')^2+(z+h)^2] + \ln[(y-y')^2+(z-h)^2 \right] \delta n_t(y').
\end{equation}
The other necessary equations are the continuity equation
\begin{equation}
%B3
e \frac{\partial \delta n_t(y)}{\partial t} + \frac{\partial j^{\ss \bot}_t(y)}{\partial y}  =0,
\end{equation}
which relates the density to the transverse current, and the equation
\begin{equation}
%B4
\delta n_t(y) = e \rho_{\ss 2D}[v_t(y)- \varphi_t(y)],
\end{equation}
where $\varphi_t(y)=\Phi_t(y,-h)$ is the electrostatic potential in the
2D plane and $v_t(y)$ is the electrochemical potential. Finally, there is
a relation between the transverse current and electrochemical potential,
\begin{equation}
%B5
j_t^{\ss \bot}(y)=\int_0^t dt'\left( -\sigma_{tt'}^{\ss\|} \frac{\partial}{\partial y} v_{t'}(y)+
\sigma_{tt'}^{\ss \perp} E^{\ss\Vert }\right),
\end{equation}
which is a generalization of Eq. (9): $E^{\ss \bot}_{t'}$ replaced by
$-\partial v_{t'}(y)/\partial y$ in order to take into account
the spatial inhomogeneity.

Using Eqs. (B2), (B3), and (B4), one can exclude the electrostatic potential
and electron density, and write
\begin{equation}
%B6
-\frac{\partial j_t^{\ss \bot}(y)}{\partial y} + \frac{2e^2 \rho_{\ss 2D}}{\epsilon}
\int_{-d/2}^{d/2} d y' K(y,y') \frac{\partial j_t^{\ss \bot}(y')}{\partial y'} =
e^2 \rho_{\ss 2D} \frac{\partial v_t(y)}{\partial t},
\end{equation}
where $K(y,y')=\ln|y-y'| + \ln \sqrt{(y-y')^2+(2 h)^2}$. Equations (B5) and
(B6), with the boundary condition $j_t^{\ss \bot}(\pm d/2)=0$, give a complete
description of the spatial distribution of transverse current and electrochemical
potential in the long Hall bar. The spatial inhomogeneity is essential near the
edges $\pm d/2$, this follows from the fact that the transverse current goes
to zero at $y=\pm d/2$. On the other hand, in the main part of the bar one
should expect nearly homogeneous currents and fields. If the bar width $d$
is large in comparison to the Bohr radius $\hbar^2 \epsilon/e^2 m$, the
main contribution to the right-hand side of Eq. (B6) comes from the integral
term, which is written after integration by parts as $(2e^2 \rho_{\ss 2D}/\epsilon)
\int_{-d/2}^{d/2} d y' [\partial  K(y,y') / \partial y] j_t^{\ss \bot}(y')$.
Let us search for the solution of Eq. (B6) in the form $j_t^{\ss \bot}(y)=
j_t^{\ss \bot}+\delta j_t(y)$, where $j_t^{\ss \bot}$ is the
homogeneous part of the current and $\delta j_t(y)$ is the
inhomogeneous correction. Neglecting first $\delta j_t(y)$ (zero-order
iteration), we obtain $j_t^{\ss \bot}= \epsilon \{2 [K(y,-d/2)-K(y,d/2)] \}^{-1}
[\partial v_t(y)/\partial t]$, which is rewritten, under the reasonable
assumption $d \gg 4h$, as
\begin{equation}
%B7
j_t^{\ss \bot}=
\frac{\epsilon}{4 \ln|(d/2+y)/(d/2-y)| } \frac{\partial v_t(y)}{\partial t}.
\end{equation}
Generally speaking, this equation can be satisfied for all $y$ if the transverse
field is inhomogeneous. In other words, $v_t(y)=-E^{\ss \bot}_t y + \delta v_t(y)$,
where $\delta v_t(y)$ is the inhomogeneous correction to the electrochemical potential.
However, in the central part of the sample, where $\ln|(d/2+y)/(d/2-y)| \simeq
4y/d + O[(y/d)^3]$, the inhomogeneous correction $\delta v_t(y)$ can be
neglected and the electrochemical potential is expressed through the homogeneous
transverse field, $v_t(y) \simeq -E^{\ss \bot}_t y$. Equation (B7) in these conditions is
reduced to Eq. (11) with the effective capacitance $C_{\ss \bot}=\epsilon d/16$ 
(so the coefficient $\alpha$ introduced after Eq. (11) is equal to 1/16), while
Eq. (B5) is reduced to Eq. (9). Using these solutions, one can write an equation
for the inhomogeneous correction to the current (first-order iteration), which
shows that the relative contribution of $\delta j_t(y)$ is small, $d^{-1}
\int_{-d/2}^{d/2} d y  \delta j_t(y) \ll j_t^{\ss \bot}$, this smallness is
of numerical origin. Therefore, the homogeneous approximation described by
Eqs. (9) and (11) is justified.


\begin{thebibliography}{16}
\bibitem{12} It is interesting to note that a complete neglect of the retardation
of edge charging, formally corresponding to $\Omega \rightarrow \infty$, would
lead to a hyperbolic singularity in the temporal dependence of the Hall field.
The analytical solution of Eq. (14) in this case is $E^{\ss \perp}_{t}/E^{\ss \|}=-
(\omega_c/g_t) \int_0^t dt' e^{-\nu(t-t')} g_{t'}$, and the Hall field
diverges when the factor $g_t$ changes its sign from negative to positive.
\end{thebibliography}
\end{document}